\documentclass[pra,twocolumn,,superscriptaddress,a4paper]{revtex4}
\usepackage{graphicx,amsmath,bm}

\begin{document}

\title{Loophole-free Bell test for one atom and less than one photon}
\date{\today}
\author{N. Sangouard}
\affiliation{Group of Applied Physics, University of Geneva, CH-1211 Geneva 4, Switzerland}
\author{J.-D. Bancal}
\affiliation{Group of Applied Physics, University of Geneva, CH-1211 Geneva 4, Switzerland}
\author{N. Gisin}
\affiliation{Group of Applied Physics, University of Geneva, CH-1211 Geneva 4, Switzerland}
\author{W. Rosenfeld}
\affiliation{Fakultat fur Physik, Ludwig-Maximilians-Universitat Munchen, D-80799 Munchen, Germany}
\author{P. Sekatski}
\affiliation{Group of Applied Physics, University of Geneva, CH-1211 Geneva 4, Switzerland}
\author{M. Weber}
\affiliation{Fakultat fur Physik, Ludwig-Maximilians-Universitat Munchen, D-80799 Munchen, Germany}
\author{H. Weinfurter}
\affiliation{Fakultat fur Physik, Ludwig-Maximilians-Universitat Munchen, D-80799 Munchen, Germany}

\begin{abstract}
We consider the entanglement between two internal states of a single atom and two photon number states describing either the vaccum or a single photon and thus containing, on average, less than one photon. We show that this intriguing entanglement can be characterized through substantial violations of a Bell inequality by performing homodyne detections on the optical mode. We present the experimental challenges that need to be overcome to pave the way towards a loophole-free Bell test in this setup.
\end{abstract}
\maketitle


\section{Introduction}

Is quantum physics a complete theory or does the description of Nature's laws require local hidden variable theories? The answer to this question, which has been asked by Einstein, Podolsky and Rosen in 1935, can be found by realizing a Bell test \cite{Bell64}. On the one hand, two distant observers who performed appropriate measurements on entangled photon pairs, have observed correlated results violating a Bell inequality, even though the measurement choices were made long after the pair creation \cite{Tittel98} and even though the photons were too far from each other to agree on the results once they knew the measurement basis \cite{Locality}. On the other hand, two ions close to each other, have also exhibited the violation of a Bell inequality even though they were forced to give a result at each trial \cite{Rowe01}. But to constitute a definitive answer, it would be necessary to close all the loopholes in the same Bell experiment, i.e. to perform a Bell test both at a distance and with high detection efficiencies.\\  

Closing the detection loophole for the Clauser-Horne-Shimony-Holt (CHSH) inequality \cite{Clauser69} requires overall detection efficiencies larger than 82.8\% for a maximally entangled state and larger than 66.7\% using partially entangled states \cite{Eberhard93} in the absence of other imperfections. This threshold detection efficiency can further be lowered using states with a dimension higher than qubits. For example, in Ref. \cite{Vertesi10}, it has been shown that a detection efficiency of 61.8\% can be tolerated using four dimensional states and a four-setting Bell inequality. However, considering realistic noise and achievable coupling into the quantum channel (usually an optical fiber) and detection efficiencies, one rapidly becomes aware that closing the detection loophole in an optical Bell test is extremely challenging. But let us keep hope alive! \\
The problem of the single-photon detection efficiency might be circumvented by using homodyne measurements which are known to be very efficient \cite{Nha04, Garcia04, Ji10}.  In this framework, theoretical proposals leading to substantial violations of Bell's inequalities and combining feasible states and measurements have been put forth recently \cite{Cavalcanti10}. \\
An attractive alternative is to use an asymmetric configuration involving e.g. atom-photon entanglement. Since the atom can be detected with an efficiency close to one, the detection efficiency on the photon side is lower than the case where the detections at both sides are inefficient \cite{Cabello07, Brunner07}, as low as 50\% for the CHSH inequality and 43\% using a three-setting inequality. Furthermore, the photon is naturally used to distribute entanglement over long distances so that the choice of the measurement on one side and the measurement result on the other side can easily be spacelike separated. Note that the entanglement between internal states of an atom and the polarization degree of freedom of a photon have already been observed experimentally \cite{Moehring04, Volz06, Rosenfeld08}. Such entanglement has further been used to entangle remote atoms from an entanglement swapping operation \cite{Moehring07}. We focus on the entanglement between internal states of an atom and a partially filled optical mode, containing on average less than one photon, as described in detail in section \ref{sec1}. We propose Bell type scenarios either combining a homodyne detection and a photon counting on the optical mode or using homodyne detections only to characterize this special entanglement. Although homodyne detections are used, we show in section \ref{sec2} that unexpectedly large violations of the CHSH inequality could be observed. We also present a feasibility study in section \ref{sec3}. We provide the minimal entanglement generation and photon counting efficiencies that are required to close the detection loophole. We then give the typical distance that is necessary to close the locality loophole. We also take the branching ratios into account, we analyze the effect of the atomic motion and we present the requirement on the optical path length stability. The last section is devoted to the conclusion.\\


\section{Entanglement creation between one atom and less than one photon}
\label{sec1}
\begin{figure}[ht!]
\includegraphics[width=5 cm]{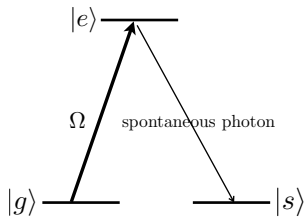}
\caption{Basic level scheme for the creation of entanglement between one atom and one optical mode containing on average less than one photon. The branching ratio is such that when the atom is excited, it decays preferentially in $s.$}
\label{fig1}
\end{figure}
Let us start by a description of the methods enabling the creation of entanglement between two atomic states and a single optical mode containing on average less than one photon. Consider an atom with a lambda-type level configuration (as depicted in Fig. \ref{fig1}), initially prepared in the state $g.$ A pump laser pulse with the Rabi frequency $\Omega$ partially excites the atom in such a way that it can spontaneously decay into the level $s$ by emitting a photon \cite{commentduration}. Long after the decay time of the atom, the atom-photon state is given by 
\begin{equation}
\label{atom_halfphoton_ent}
\psi^{\phi}=\cos{\theta} |g,0\rangle + e^{i\phi} \sin{\theta} |s,1\rangle
\end{equation}
where $\!\theta= \frac{1}{2}\int ds$  $\Omega(s)$ refers to the area of the pump pulse. The phase term is defined by $\phi=k_p r_p-k_s r_s$ where $k_p$ ($k_s$) corresponds to the wave vector of the pump (the spontaneous photon) and $r_p$ ($r_s$) is the atom position when the pump photon is absorbed (the spontaneous photon is emitted). Note that $\phi$ may vary in practice, e.g. due to atom position variations. The requirements for the phase stability are studied in detail below but we first answer the question: can the entanglement between an atom and a partially filled optical mode be measured from the violation of a Bell inequality? \\


\section{Homodyne detections in an asymmetric Bell test}
\label{sec2}

\subsection{Principle of the Bell test}

First, let us recall the principle of a Bell-CHSH test. Two distant observers, usually named Alice and Bob, share a quantum state. Each of them chooses randomly a measurement among two projectors, $\{X_i\}$ for Alice, $\{Y_j\}$ for Bob, $i,j \in [1,2]$ and obtains a binary result, $\{a_i\}$ and $\{b_j\}$ for Alice and Bob respectively. By repeating the experiment several times, Alice and Bob can compute the conditional probabilities $p(a_ib_j|X_iY_j).$ They can then easily deduce the value of the CHSH parameter
\begin{equation}
\label{Bell_inequality}
S=E_{X_1Y_1}+E_{X_1Y_2}+E_{X_2Y_1}-E_{X_2Y_2}
\end{equation}
where $E(X_iY_j)=p(a_i=b_j|X_iY_j)-p(a_i \neq b_j|X_iY_j).$ Alice and Bob will conclude that the observed correlations cannot be described by local hidden variable theories if they find measurement settings such that $S > 2.$ Note that all possible states leading to a violation of a Bell inequality are entangled. Therefore, a Bell test can be seen as a test of the laws of Nature but also as a witness of entanglement.\\

\subsection{Bell test with one atom and less than one photon}

Now, consider the specific case where Alice and Bob share a state of the form (\ref{atom_halfphoton_ent}). Alice applies projective measurements on the atomic states and can freely choose projections on arbitrary vectors $\overrightarrow{v_j}=\cos \frac{\alpha_j}{2} |g\rangle+e^{i\varphi_j} \sin \frac{\alpha_j}{2} |s\rangle$ of the Bloch sphere. For each measurement $X_j,$ $j=1,2$, Alice sets $a_j=+1$ if she gets a result along $\overrightarrow{v_j}$ and $a_j=-1$ if the result is directed along $\overrightarrow{v_j}^{\bot}.$ Bob applies measurements on the optical mode and chooses either to count the photon number $Y_1=n$ or to measure the quadrature $Y_2=\cos \zeta \hat{X}+\sin \zeta \hat{P}.$ When he measures $n,$ he naturally sets the results $b_1=+1$ if the result is positive and $b_1=-1$ if there is no photon. When he performs the quadrature measurement, he gets a real number $x.$ He then has to process this result to get binary outcomes. He decides to attribute the results $b_2=-1$ if the result is negative $x \leq 0$ and $b_2=+1$ otherwise. \\

We now show that Alice and Bob can obtain a substantial violation of the CHSH inequality for appropriate settings. But let us first detail the calculation of probability distributions $p(a_ib_j|X_iY_j)$ for the four pairs of measurements separately. When Bob measures $n,$ he gets $b_1=-1$ with the probability $\cos^2 \theta$ and Alice's qubit is projected into $|g\rangle.$ Therefore, $$p(+1 ,-1|X_jY_1)=\cos^2 \theta |\langle \overrightarrow{v_j} | g \rangle|^2 = \cos^2 \theta \left(\frac{1+ \cos \alpha_j}{2}\right).$$ Similarly $$p(-1, -1|X_jY_1)=\cos^2 \theta |\langle \overrightarrow{v_j}^{\bot} | g \rangle|^2 = \cos^2 \theta \left(\frac{1- \cos \alpha_j}{2}\right).$$  Following similar lines for $b_1=+1,$ one finds $$p(a_j ,+1|X_j Y_1)=\sin^2 \theta \left(\frac{1-a_j \cos \alpha_j}{2}\right)$$ leading to 
\begin{equation}
E_{X_jY_1}=-\cos \alpha_j.
\end{equation}
When Bob measures $Y_2$ and obtains $b_2=-1,$ Alice's state is projected into
\begin{eqnarray}
\nonumber 
&&\rho^A_{b_2=-1}=\cos^2 \theta \int_{-\infty}^{0} dx |\Phi_0 (x)|^2 |g\rangle\langle g|\\
\nonumber
&& + \frac{1}{2} \sin 2\theta e^{-i\phi} e^{i\zeta} \int_{-\infty}^{0} dx \Phi_1^{\star} (x)\Phi_0 (x) |g\rangle\langle s|\\\nonumber
&& + \frac{1}{2} \sin 2\theta e^{i\phi} e^{-i\zeta} \int_{-\infty}^{0} dx \Phi_0^{\star} (x)\Phi_1 (x) |s\rangle\langle g|\\
\nonumber 
&& + \sin^2 \theta \int_{-\infty}^{0} dx |\Phi_1 (x)|^2 |s\rangle\langle s|
\end{eqnarray} 
where $\Phi_0 (x)=\langle x | 0\rangle$ and $\Phi_1 (x)=\langle x | 1\rangle$ are the probability amplitude distribution for the vacuum and single photon Fock states ($\Phi_n(x)=\frac{1}{(2^n n! \sqrt{\pi})^{1/2}}H_n(x)e^{-x^2/2}$ where $H_n(x)$ is the Hermite polynomial). $p(+1, -1|X_j Y_2)$ ($p(-1 ,-1|X_j Y_2)$) is merely deduced from $\langle \overrightarrow v_j | \rho^A_{b_2=-1} | \overrightarrow v_j \rangle$ ($\langle \overrightarrow v_j^{\bot} | \rho^A_{b_2=-1} | \overrightarrow v_j^{\bot} \rangle$). One can check that $p(a_j ,+1|X_j Y_2)$ has the same expression than $p(a_j, -1|X_j Y_2)$ but where the integration over dx runs from 0 to $+\infty.$ One finds
\begin{equation}
\nonumber
E_{X_jY_2} =\sqrt{\frac{2}{\pi}} \sin \alpha_i \sin 2\theta \cos(\varphi_j-\phi+\zeta).
\end{equation}
Interestingly, this expression is the same, up to a factor of $\sqrt{2/\pi}$, as the expression of the correlator when Bob applies a perfect qubit measurement along $\cos\zeta\sigma_x+\sin\zeta\sigma_y$. This invites us to interpret the homodyne measurement above (with the binning $x \leq 0 \rightarrow b_2=-1$, $x > 0 \rightarrow b_2=+1$ and in the $\{|0\rangle, |1\rangle\}$ subspace), as a noisy qubit measurement in the x-y plane of the Bloch sphere with visibility $\sqrt{2/\pi}$, as was also noticed in \cite{Quintino11}.\\
Substituting the correlators by their expressions into (\ref{Bell_inequality}), one obtains a value of the CHSH polynomial for any state of the form (\ref{atom_halfphoton_ent}) and for any measurement of Alice. We found the maximal violation $S=-2\cos\alpha_1+2\sqrt{2/\pi}\sin\alpha_1\approx2.56$ for $\theta=\pi/4,$ $\phi=0$ i.e. the $\phi_+$ state and for $\varphi_1=\varphi_2=\zeta=0,$ and $\alpha_1=-\alpha_2=2\arctan(\frac{\sqrt{\pi}+\sqrt{2+\pi}}{\sqrt{2}}).$ This violation is the largest that we know in a scenario involving a homodyne detection where both the measurements and the state could be realized experimentally (see \cite{Cavalcanti10} and references therein). \\

\subsection{Bell test with homodyne detections only on the optical mode}

A natural question is whether a violation of the CHSH inequality can also be observed by measuring the optical mode with homodyne detections only. It turns out that a violation $S = 4/\sqrt{\pi} \approx 2.26$ can indeed be obtained if Alice and Bob share a maximally entangled state of the form (\ref{atom_halfphoton_ent}) with $\theta=\pi/4$ and $\phi=0$, provided that Bob's measurements are performed in complementary quadratures $Y_1=\hat X$ and $Y_2=\hat P$ and that Alice's measurements correspond to projections along vectors spanning the (xy) plane with angles $\pm 45 \deg$ between them. This result can easily be understood using the analogy previously mentioned.
If Bob would have used either $\sigma_x$ or $\sigma_y,$ the CHSH parameter would have been saturated $S=2\sqrt{2}.$ Since $\hat X$ and $\hat P$ correspond to such measurements but with the reduced visibility $\sqrt{2/\pi},$ S is reduced by the corresponding factor. \\


\section{Imperfections} 
\label{sec3}
So far, we have shown that an ideal realization would lead to significant violations of the CHSH inequality. However, the story would not be complete without a discussion taking experimental imperfections into account.\\


\subsection{Transmission inefficiency}
\begin{figure}[ht!]
\includegraphics[width=9 cm]{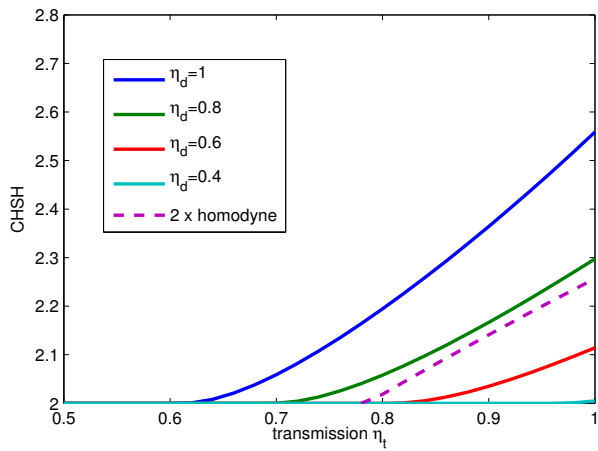}
\caption{Robustness of the CHSH violation with respect to the transmission efficiency $\eta_t.$ The full lines correspond to the case where Bob chooses either a photon counter or a measurement of a field quadrature. The upper (blue) curve is associated to a photon detector with unit efficiency $\eta_d=1$. The three other full lines are associated to inefficient counting (from $\eta_d=0.8$ to $0.4).$ The dashed line is associated to the case where Bob uses two homodyne detections (with unit efficiencies).}
\label{fig2}
\end{figure}

Let $\eta_t$ be the transmission efficiency which accounts for all the coupling inefficiencies from the atom to Bob's location. The probability amplitude associated to $|s,1\rangle$ is now multiplied by $\sqrt{\eta_t}$ and Alice and Bob can share the state 
$
\psi^{\phi}_{\eta_t}=\frac{1}{\sqrt{N}}\left(\cos{\theta} |g,0\rangle + e^{i\phi} \sin{\theta} \sqrt{\eta_t} |s,1\rangle\right)
$
with the probability $N=\cos{\theta}^2+\sin{\theta}^2 \eta_t.$ Alternatively, the photon can be lost. Tracing out the lost photon, the resulting state is $|s,0\rangle$ and it contributes to the global state with a weight $\sin^2\theta \left(1-\eta_t\right).$ To know the sensitivity of the CHSH inequality with respect to the transmission inefficiency, we thus have to compute S from the overall state
\begin{equation}
\rho_{\eta_t}={N} |\psi^{\phi}_{\eta_t}\rangle \langle \psi^{\phi}_{\eta_t}|+\sin^2\theta \left(1-\eta_t \right) |s,0\rangle\langle s,0|
\end{equation} 
for all possible values of $\theta, \phi, \varphi_1,\varphi_2, \zeta, \alpha_1$ and $\alpha_2$ as a function of $\eta_t.$ The result is shown in Fig. \ref{fig2}. In the scenario where Bob uses a photon counter and a homodyne detection with unit efficiencies, a transmission efficiency of $\eta_t=61\%$ can be tolerated (see blue curve). Although this is certainly demanding, recent results suggest that this might soon be within reach of experiments \cite{Mundt02}. It is also interesting to study the sensitivity of the violation with respect to the detection inefficiency. The homodyne measurements can fairly be considered to have unit efficiencies but most of the single-photon detectors are inefficient. Let $\eta_d$ be the efficiency of the photon counting detector. From an optimization similar to the previous one, we found that a threshold detection of $\eta_d=39\%$ can be tolerated for a transmission with unit efficiency. Our scheme is less sensitive to counting inefficiency than transmission inefficiency since the former affects only one of Bob's measurements. The two previous efficiency thresholds can be compared with a scheme that exhibits the same asymmetry but uses the entanglement with the polarization mode \cite{Brunner07} and where the violation of the CHSH inequality requires $\eta_t\eta_d \geq 50\%$. (The effect of detection and transmission imperfections is the same in this case, since Bob uses two photon counting detectors.) The latter is less sensitive to inefficiency in the transmission (for ideal detectors with $\eta_d=1$), while the scheme we propose is less sensitive to the detector inefficiency (for transmission with $\eta_t=1.$) Note also that regarding the results presented in Refs. \cite{Brunner07, Vertesi10} where the threshold efficiency has been lowered using inequalities with more settings, one could have hoped improvements using inequalities different from the CHSH inequality. However, we could not find better resistances with additional binnings for Bob's results and for more (up to three) inputs.\\

In the case where Bob uses only quadrature measurements, the CHSH inequality can be violated provided that the transmission efficiency is larger than 79\%. This result cannot fairly be compared with other schemes using atom-photon entanglement since this is the only one that we know where all the detections can be reasonably considered perfectly efficient and where only the transmission inefficiency decreases the violation.\\

\subsection{Required distance between Alice and Bob} In the previous subsection, we have addressed efficiency issues related with the photon detection and with the transmission. If we intend to close the locality loophole too, we have to give an answer how long the state detection takes. It is likely reasonable to believe that the detection time is limited by the atom \cite{note_detectime}. If the atomic states are readout on the basis of stimulated Raman adiabatic passage, ultrafast laser-ionization and registration of the correlated electron-ion pairs with coincident counting via two opposing channel electron multipliers \cite{Henkel10}, we can reach a measurement time of less than 1 $\mu s.$ Therefore, the locality loophole could be closed if Alice and Bob are separated by 300 m. For 800nm photons, the losses are of 2dB/km. This would translate into a transmission of 93\%.


\subsection{Branching ratio} So far, we have considered that once the atom is excited, it decays into the state $s.$ Consider the more realistic case where the decay from $e$ to $s$ occurs with the probability $f_s.$ Let $f_g$ be the probability for a decay into $g$ and $f_{\text{aux}}$ the decay probability into other auxiliary states such that $f_s+f_g+f_{\text{aux}}=1.$ Taking these branching ratios into account, the state long after the interaction with the pump pulse is 
\begin{eqnarray}
\nonumber
& \rho_f &={N'} |\psi^{\phi}_{f_s}\rangle \langle \psi^{\phi}_{f_s}|+\sin^2\theta f_g |g,0\rangle\langle g,0|\\
&&+\sin^2\theta f_{\text{aux}} |\text{aux},0\rangle\langle \text{aux},0|.
\end{eqnarray}
$\psi^{\phi}_{f_s}$ is defined from $\psi^{\phi}_{\eta_t}$ where $\eta_t$ is replaced by $f_s$ and $N'=\cos{\theta}^2+\sin{\theta}^2 f_s.$ \\
We now present a strategy to make $S>2$ (calculated from this state) as soon as $f_s \neq 0.$ If Alice chooses to attribute the result $a_j=-1$ when she measures the atom in the state $\text{aux},$ the correlators calculated from $|\text{aux},0\rangle$ are $E_{X_jY1}=1, E_{X_jY2}=0 \-\ \forall j \in \{1,2\}$ and the resulting S value is equal to $2$ in this case. Moreover, if she chooses two measurements very close to $\sigma_z,$ i.e. $(\alpha_1=\pi-\epsilon,\varphi_1=0)$ and $(\alpha_2=-\pi+\epsilon,\varphi_2=0),$ one can check that the S value computed from the component $ |g,0\rangle$ is $2-\epsilon^2.$ If she further excites the atom so that $\theta=\pi/4$ and $\phi=0,$ and if Bob chooses $\zeta=0,$ the S value from $\psi^{\phi}_{f_s}$ is roughly $2+4\frac{\epsilon}{1+f_s} \frac{\sqrt{2f_s}}{\sqrt{\pi}}.$ The overall CHSH value is thus given by 
\begin{equation}
S \approx 2+4\frac{\sqrt{f_s}}{\sqrt{2\pi}} \epsilon+o(\epsilon^2).
\end{equation}
Therefore, in the absence of errors, a violation of the CHSH inequality can be observed as soon as the probability that the excited atom decays into $s$ is non-zero. This Bell test can thus be applied to a large number of atomic species since it is very resistant to branching ratio variations. However, the larger the decay into $s$ is, the larger the violation. \\


\subsection{Stability requirements}
Let us now focus on the phase stability constrains. The phase term of the state (\ref{atom_halfphoton_ent})
has to stay stable from trial to trial. In practice, however, $\phi$ may vary, e.g. due to atom position variations. For wavevectors with the same norm $||\overrightarrow k||=||\overrightarrow {k_p}||=||\overrightarrow {k_s}||,$ the fidelity of the resulting entanglement 
\begin{equation}
\rho=F|\psi^\phi \rangle\langle \psi^\phi |+(1-F)|\psi^{\phi+\pi}\rangle\langle\psi^{\phi+\pi}|
\end{equation}
is found to be  
\begin{equation}
F=\frac{1}{2}\left(1+e^{-2a^2(\bar n+1/2)\Delta k}\right)
\end{equation}
in the weak confinement regime \cite{Cabrillo99, Luo09}. $a=\sqrt{\hbar/(2 m \omega)}$ is the size of the harmonic trapping potential ground state for an atom of mass $m$ within a trap of frequency $\omega.$ $\bar n$ is the average number of thermal quanta of motion and $\Delta k = ||\overrightarrow k|| (1-\cos{\theta})$ where $\theta$ is the angle between the pump beam and the emission direction. Hence, the problem of the atomic motion can not only be overcome by cooling the ions deeply within the Lamb-Dicke limit (where $\bar n$ is small) but more simply by collecting the photons scattered in the forward direction where $\theta=0.$ \\

Let us also comment the stability requirement on the optical path lengths. The local oscillator which is required to perform the homodyne detections at Bob's location could be obtained by picking off a fraction of the pump beam with a beamsplitter. In this case, the setup will be made of a large Mach-Zehnder interferometer and the path length difference between the two arms of the interferometer $\Delta L$ has to be stable so that $||\overrightarrow k|| \Delta L \ll 1.$ Note that temperature variations change both the refraction index (and thus $||\overrightarrow k||$) and the length of fibers $\Delta L$. For several tens of kilometers long commercial fibers installed in an urban environment, the typical time needed for a mean phase change of 0.1 rad (corresponding to a fidelity of 0.9) is of the order of 100 $\mu$s \cite{Minar08}. This lets us believe that an active stabilization of the phase should be possible even for very long interferometers using available technologies. This is well confirmed by recent experimental results \cite{Cho09}.\\

\section{Conclusion}

We proposed different scenarios to measure the entanglement between the internal state of an atom and an optical mode containing, on average, less than one photon. We reported large violations of a CHSH inequality for both the case where one homodyne detection and one photon counting are performed on the optical mode and for two homodyne detections. With homodyne detections only, a minimal entanglement generation efficiency of 79\% can be tolerated. This efficiency goes down to 61\% if homodyne detections are combined with unit efficiency photon counting. We have also shown that in principle, a violation of the CHSH inequality can be obtained even for branching ratios favoring the photon emission in undesired modes. There is no need to cool the atom deeply within the Lamb-Dicke regime if the scattered photons are collected close to the forward direction with respect to the pump propagation. Finally, the stability requirements for the optical path lengths is within reach of experiments. \\

We believe that our work could provide motivations for several research groups. A lot of efforts have already been devoted into the characterization of single-photon Fock states with homodyne detections \cite{single_ph_homo}. Moreover, although the setup recently developed by the Rempe group has been used to address squeezed light \cite{Ourjoumtsev11}, it is of particular interest for our proposal as it combines a single atom embedded a high finesse cavity with a homodyne detection. Note also that, behind its fundamental interest, our proposal might find exciting applications in the framework of quantum information sciences, e.g. for device-independant quantum cryptography \cite{diqkd}.\\

We thank M. Afzelius, N. Brunner, D. Cavalcanti, H. De Riedmatten, R. Thew for interesting discussions. We gratefully acknowledge support by the EU project Qessence, from the EU ERC AG Qore and the Swiss NCCRs QP and QSIT.


\begin{thebibliography}{3}
\bibitem{commentduration} Note that the pump pulse duration is considered to be short relative to the lifetime of the excited state in order to minimize two-photon emissions.
\end{thebibliography}
\end{document}